\documentclass[runningheads,a4paper]{llncs}

\usepackage{amssymb}
\usepackage{amsmath}
\usepackage{graphicx}

\usepackage{epsfig}
\usepackage{graphicx}
\usepackage{dcolumn}
\usepackage{bm}
\usepackage{tikz}
\usepackage[all]{xy}
\usetikzlibrary{arrows,shapes}
\usepackage{url}
\usepackage{centernot}
\usepackage{stmaryrd}

\begin{document}

\mainmatter  

\title{A Markov model for inferring flows in directed contact networks
}

\titlerunning{A Markov model for inferring flows in directed contact networks}

%
%
\author{Steve Huntsman
}
\authorrunning{S. Huntsman}

\institute{
BAE Systems FAST Labs, 4301 North Fairfax Drive, Arlington, Virginia 22203, USA\\
\email{steve.huntsman@baesystems.com}
}

%
%

\toctitle{A Markov model for inferring flows in directed contact networks}
\tocauthor{S. Huntsman}
\maketitle

\begin{abstract}
\emph{Directed contact networks} (DCNs) are a particularly flexible and convenient class of temporal networks, useful for modeling and analyzing the transfer of discrete quantities in communications, transportation, epidemiology, etc. Transfers modeled by contacts typically underlie \emph{flows} that associate multiple contacts based on their spatiotemporal relationships. To infer these flows, we introduce a simple inhomogeneous Markov model associated to a DCN and show how it can be effectively used for data reduction and anomaly detection through an example of kernel-level information transfers within a computer.
\end{abstract}

\section{\label{sec:Introduction}Introduction}

To identify and prevent data exfiltration by advanced persistent threats (APTs), it is useful to analyze graphs that encode causality data about system activities. The form of this causality data varies widely among providers, but it can be cariacatured as tagged and tracked information (micro-)transfers. In practice, these are realized as annotations of system calls or similar kernel-level events \cite{ChanEtAl,JenkinsonEtAl}. We have found the following intermediate representation or \emph{summary} useful:
\begin{equation}
(\text{\emph{timestamp}},\text{\emph{subject}},\text{\emph{verb}},\text{\emph{object}}). \nonumber
\end{equation} 
In our tools, this takes the explicit form 
\begin{equation}
\label{eq:summary}
(\text{\emph{timestamp}},(\text{\emph{process name}},\text{\emph{process ID}}),\text{\emph{event type}},\text{\emph{filename}})
\end{equation} 
where the ``everything is a file'' philosophy applies to the last entry (e.g., for a fork event, the filename is the forked process ID).

Meanwhile, any causal representation of the transfer of some token must involve a \emph{source} $s$, a \emph{target} $t$, and some notion of the \emph{time} $\tau$ at or over which the transfer occurs. The simplest notion of such a time is a single instant, and the resulting notion of a \emph{(directed) contact} as an ordered triple $(s,t,\tau)$ is not only the simplest, but also a very general causal representation of a transfer. For example, a transfer from a source $s$ to a target $t$ over the time interval $[\tau_0,\tau_1]$ could be represented with the two contacts $(s,*,\tau_0)$ and $(*,t,\tau_1)$, where here $*$ is a shorthand for the triple $(s,t,[\tau_0,\tau_1])$.
\footnote{
A useful analogy is of a flight departing from $s$ at $\tau_0$ and arriving at $t$ at $\tau_1$: the contacts $(s,*,\tau_0)$ and $(*,t,\tau_1)$ respectively correspond to embarking and debarking. This analogy also highlights that alternative representations could also include additional contacts $(s,*,\tau_*)$ with $\tau_0 \le \tau_* < \tau_1$ depending on the desired behavior.
}

Many kernel-level events have an unambiguous directionality with respect to potential information transfer. For example, if process $A$ closes file $X$, this might entail deleting $X$, which can be regarded as a degenerate information transfer from $A$ to $X$ (i.e., overwriting $X$ and its metadata with $\varnothing$), but there is no possibility of information transfer from $X$ to $A$ as a result of the closure \emph{per se}. Similarly, if $A$ forks process $B$, information might be transferred from $A$ to $B$ as part of the fork, but not the other way around. In both of these examples the source $A$ can reasonably be interpreted as ``writing'' to a target in some sense. Events which can be viewed as generalized reads or writes in this way naturally correspond to directed contacts. Other events do not have an unambiguous directionality, and as such conservatively correspond to pairs of directed contacts with the source and target swapped (e.g., the act of opening a file can entail a [generalized] read or write).

This paper focuses on how replacing each event of the form \eqref{eq:summary} with either one or two contacts provides a further useful abstraction of causality/transfer data that is mathematically convenient. In particular, we show how a natural model arises for probabilistically modeling potential flows. This model involves just one parameter, and there is a simple heuristic for setting it that we follow in practice. We detail the model behavior through analytical and practical examples. It is important to note at the outset that the model is not statistical in the sense that it involves no learning, fitting, optimization, etc. Its construction instead follows the tradition of physics by starting from various required symmetries (e.g., time translation invariance) that \emph{any} model built from contacts ought to obey and deriving the most general mathematical structure that is consistent with those symmetries. As a byproduct of this generality, the model also applies to related problems that can be modeled using directed contact networks, e.g. for network traffic analysis or disease surveillance.

The difference between \eqref{eq:summary} and a directed contact is manifested in two graphs that we have used to analyze system behavior. The \emph{enterprise provenance graph} (EPG) is morally an undirected multigraph, with vertices labeled by files (including processes) and edges labeled by event types and timestamps.
\footnote{
In fact the EPG is a directed graph with vertices bipartitioned into files and events; arcs indicating a subject or object go from events to files. However, there is an obvious bijective correspondence between this and our moral characterization.
}
The \emph{temporal digraph} (TD) is directed, with vertices labeled by ordered pairs of files and timestamps, \emph{spatial arcs} corresponding to contacts as outlined above, and \emph{temporal arcs} linking files through time. While the EPG has demonstrated its utility in several forensic challenges under the DARPA Transparent Computing program and supports time-aware backtracking \cite{KingChen}, it represents the passage of both time and (potentially) information implicitly through annotations, sacrificing explicit structure for precision and expressiveness. The TD places this same structure above other considerations (though its arcs could also easily be annotated with summarized events). This has several distinct benefits: for example, it turns the backtracking problem into a trivial breadth-first search, and also naturally leads to the model at the heart of the present paper.

Notwithstanding the context of information transfer, the closest work to ours is \cite{SerGiacomiEtAl}, which demonstrates that the most probable paths in a Markovian model of a very complicated temporal network (viz., ocean water transport in the Mediterranean) suffice to describe the network's key features. The particulars of our model and context are very different apart from the gross feature of Markovity, but our conclusion is essentially identical: the most probable paths/flows suffice for capturing the key information dynamics. This particularly includes highly probable but nevertheless infrequent (or one-time) flows that reliably capture anomalous or even malicious behavior.

The paper is structured as follows: \S \ref{sec:DirectedContactNetworksAndTemporalDigraphs} introduces directed contact networks and temporal digraphs; \S \ref{sec:Markov} discusses our Markov model; \S \ref{sec:DRAD} discusses its performance in data reduction and anomaly detection, and \S \ref{sec:Remarks} concludes the paper.

\section{\label{sec:DirectedContactNetworksAndTemporalDigraphs}Directed contact networks and temporal digraphs}

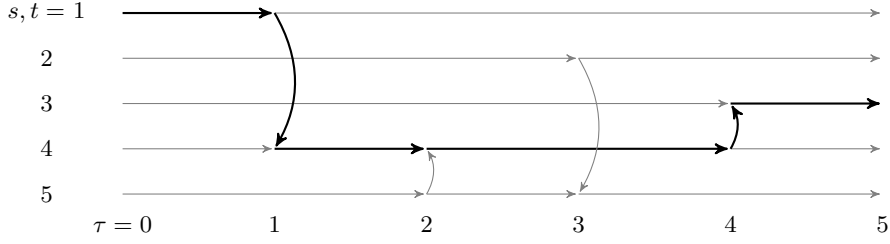
\begin{figure}
	\begin{tikzpicture}[->,>=stealth',shorten >=1pt]
		\node at (-1,-0.6) {$s,t = $ 1};
		\node at (-1,-1.2) {2};
		\node at (-1,-1.8) {3};
		\node at (-1,-2.4) {4};
		\node at (-1,-3.0) {5};
		\node at (0,-3.4) {$\tau = 0$};
		\node at (2,-3.4) {$1$};
		\node at (4,-3.4) {$2$};
		\node at (6,-3.4) {$3$};
		\node at (8,-3.4) {$4$};
		\node at (10,-3.4) {$5$};
		\coordinate (v1a) at (0,-0.6);
		\coordinate (v2a) at (0,-1.2);
		\coordinate (v3a) at (0,-1.8);
		\coordinate (v4a) at (0,-2.4);
		\coordinate (v5a) at (0,-3.0);
		\coordinate (v11) at (2,-0.6);
		\coordinate (v21) at (2,-1.2);
		\coordinate (v31) at (2,-1.8);
		\coordinate (v41) at (2,-2.4);
		\coordinate (v51) at (2,-3.0);
		\coordinate (v42) at (4,-2.4);
		\coordinate (v52) at (4,-3.0);
		\coordinate (v23) at (6,-1.2);
		\coordinate (v53) at (6,-3.0);
		\coordinate (v14) at (8,-0.6);
		\coordinate (v24) at (8,-1.2);
		\coordinate (v34) at (8,-1.8);
		\coordinate (v44) at (8,-2.4);
		\coordinate (v54) at (8,-3.0);
		\coordinate (v1z) at (10,-0.6);
		\coordinate (v2z) at (10,-1.2);
		\coordinate (v3z) at (10,-1.8);
		\coordinate (v4z) at (10,-2.4);
		\coordinate (v5z) at (10,-3.0);
		\foreach \from/\to in {
			v11/v1z, v2a/v23, v23/v2z, v3a/v34, v4a/v41, v44/v4z, v5a/v52, v52/v53, v53/v5z}
		\draw [gray] (\from) to (\to);
		\foreach \from/\to in {
			v1a/v11, v34/v3z, v41/v42, v42/v44}
		\draw [thick] (\from) to (\to);
		\draw [thick] (v11) [out=-60,in=60,looseness=1] to (v41);
		\draw [gray] (v52) [out=60,in=-60,looseness=1] to (v42);
		\draw [gray] (v23) [out=-60,in=60,looseness=1] to (v53);
		\draw [thick] (v44) [out=60,in=-60,looseness=1] to (v34);
	\end{tikzpicture}
	\caption{\label{fig:DCNtemporalgraph} Temporal digraph of the DCN $\mathcal{C} := \{(1,4,\tau_1), (5,4,\tau_2), (2,5,\tau_3), (4,3,\tau_4)\}$ with $\tau_1 < \tau_2 < \tau_3 < \tau_4$. Note that there is a temporally coherent path from $\mathcal{C}$-vertices 1 to 3 (indicated with {\bf bold} versus {\color{gray}gray} arrows), but not from 2 to 3. Spatial (resp. temporal) arcs are vertical (resp. horizontal); temporal fibers are vertices along horizontal paths.}
\end{figure}

Digraphs admit a natural temporal generalization called \emph{directed contact networks (DCNs)} that are a particularly simple incarnation of \emph{temporal networks} \cite{Holme,MasudaLambiotte}. While we can think informally of DCNs as collections of contacts as described in \S \ref{sec:Introduction}, this allows certain degenerate situations to occur that a slightly more formal and restrictive notion will avoid. 
Towards this end, a DCN with vertex set $V \equiv [n] \equiv \{1,\dots,n\}$ is a finite nonempty set $\mathcal{C}$ for which each \emph{contact} $c \in \mathcal{C}$ corresponds to a unique triple $(s(c),t(c),\tau(c)) \in [n] \times [n] \times \mathbb{R}$ with $s(c) \ne t(c)$; when convenient we identify contacts and their corresponding triples. There is an obvious notion of a temporally coherent path which we do not bother to write out formally but which is indicated in Figure \ref{fig:DCNtemporalgraph}.

Define the \emph{temporal digraph} of $\mathcal{C}$ (see Figure \ref{fig:DCNtemporalgraph} for an example) to be the digraph $T(\mathcal{C})$ with vertex and arc sets
\begin{eqnarray}
\label{eq:TemporalDigraphVertices}
V(T(\mathcal{C})) & := & \{(v, \pm \infty) : v \in V\} \nonumber \\
& & \cup \{(v,\tau(c)) : [ (v,c) \in V \times \mathcal{C} ] \land [s(c) = v \lor t(c) = v] \} \\
\label{eq:TemporalDigraphArcs}
A(T(\mathcal{C})) & := & \{((s(c),\tau(c)),(t(c),\tau(c))) : c \in \mathcal{C} \} \nonumber \\
& & \cup \{ ((v,\tau_{j-1}^{@v}),(v,\tau_j^{@v})) : v \in V, j \in [|\mathcal{C}@v|-1] \}
\end{eqnarray} 
where the \emph{temporal fiber at $v$ is}
\begin{equation}
\label{eq:TemporalFiber}
\mathcal{C}@v := \{\pm \infty\} \cup \{\tau(c) : c \in \mathcal{C} \land (s(c) = v \lor t(c) = v) \} \equiv \{\tau_j^{@v}\}_{j = 0}^{|\mathcal{C}@v|-1}.
\end{equation}
The first set in the union on the RHS of \eqref{eq:TemporalDigraphArcs} is the set of \emph{temporal arcs}; the second set is the set of \emph{spatial arcs}. Note that $|V(T(\mathcal{C}))| = \sum_v |\mathcal{C}@v| \le 2|V| + 2|\mathcal{C}|$ and $|A(T(\mathcal{C}))| = |V(T(\mathcal{C}))| - |V| + |\mathcal{C}| \le |V| + 3 |\mathcal{C}|$, so that $T(\mathcal{C})$ can be formed with only linear overhead (though this requires some care in practice).

\section{\label{sec:Markov}Markov model associated to a DCN}

The reader may initially be skeptical that a useful probabilistic model of potential information flow can be built upon $T(\mathcal{C})$ alone. However, this claim of utility merely asserts that ubiquitous traffic-analytic products such as pen registers or trap and trace devices (defined in 18 USC \S 3127) actually produce data that allow a user to make substative inferences about the sources and diffusion of information in a communications network. In this light, the claim is obviously supported by general knowledge that informs the basic assumption of the model. 

Namely, suppose we have contacts of the form $(A,B,0)$ and $(B,C,\tau)$: the model probability that information (potentially) flows from $A$ to $C$ ought to decrease from unity to zero as $\tau \uparrow \infty$. It is important to note that this behavior does not entail that the model will have problems with capturing APTs: even a ``low and slow'' data exfiltration is almost certain to involve at least some system call-scale directed contacts that are either unusual in their own right or are temporally localized.  

Besides the qualitative requirement above, a reasonable model that assigns probabilities to the arcs of $T(\mathcal{C})$ is tightly constrained by a number of basic symmetries that it must obey, namely w.r.t.
\begin{itemize}
\item[i)] vertices (probabilities assigned to spatial arcs must not explicitly depend on their source or target);
\item[ii)] time windowing (the model must yield probabilities for information flows from a source at an initial time to a target at a terminal time that coherently compose over different sets of adjacent time windows spanning the same interval);
\item[iii)] non-local behavior (probabilities assigned to temporal arcs must only depend on their duration and the number of spatial arcs sharing the same source);
\item[iv)] simultaneous vs. infinitesimally separated events (the probabilities for these cases must only differ infinitesimally).
\end{itemize}
For the sake of brevity, we will simply construct a model that uniquely satisfies these properties, without formally interpreting them or giving proofs. However, the nature and proof of these properties should be reasonably evident to the mathematically inclined reader from the construction itself.

Define the \emph{restriction} of a DCN $\mathcal{C}$ to $X \subset \mathbb{R}$ by $\mathcal{C} |_X := \tau^{-1}(\tau(\mathcal{C}) \cap X)$, i.e., the subset of contacts with times in $X$. Next, for $a_1 \notin \tau(\mathcal{C})$ 
\footnote{
If $a_1 \in \tau(\mathcal{C})$, we can simply consider instead $a'_1 = a_1+\varepsilon_\mathcal{C}$, where $\varepsilon_\mathcal{C} := \frac{1}{2} \min_{t,t' \in \tau(\mathcal{C}), t \ne t'} |t - t'|$. Note that here we assume without loss of generality that $|\tau(\mathcal{C})| > 1$, i.e., that $\mathcal{C}$ is nontrivial as a DCN (versus, e.g., a digraph).
}
and $a_0 < a_1$, consider the \emph{restricted temporal digraph} $T(\mathcal{C})|_{[a_0,a_1)}$ obtained by modifying $T(\mathcal{C}|_{[a_0,a_1)})$ by replacing instances of $-\infty$ and $\infty$ in \eqref{eq:TemporalDigraphVertices} with $a_0$ and $a_1$, respectively, keeping the form of \eqref{eq:TemporalDigraphArcs} apart from analogous replacements in \eqref{eq:TemporalFiber}. That is, $T(\mathcal{C})|_{[a_0,a_1)}$ is obtained from $T(\mathcal{C}|_{[a_0,a_1)})$ by replacing the second component of the vertices $(v,-\infty)$ with $a_0$ and the second component of the vertices $(v,\infty)$ with $a_1$, while retaining all the arcs. 

We now introduce a physically inspired model of temporally coherent random paths in which an ``inverse temperature'' $\beta \in \mathbb{R}$ governs a balance between temporal and spatial arcs in a temporal digraph. 
\footnote{
Bearing the idea of negative absolute temperature \cite{Ramsey} in mind, we note that $\beta = -\infty$ corresponds to ``absolute hot'', and $\beta = \infty$ corresponds to absolute zero. We follow a natural convention (and it is nothing more) for our model, in which lower temperatures correspond to slower dynamics: thus $\beta \uparrow \infty$ and $\beta \downarrow -\infty$ are respectively limits in which no temporal and spatial arcs are traversed. {\bf In practice, we follow a physical analogy and set $\beta^{-1}$ to the average time between contacts.}
}
Specifically, for $\varepsilon \ge 0$, $a_0 < \dots < a_M$ with $a := \{a_m\}_{m=0}^M$, $a \cap \tau(\mathcal{C}) = \varnothing$, and $m \in [M]$, we form the Markov chain on $V(T(\mathcal{C})|_{[a_{m-1},a_m)})$ with transition matrix $ P_{(\mathcal{C},a,m)}^{(\beta,\varepsilon)}$ defined by
\begin{eqnarray}
\label{eq:transition1}
& & Z_{(v,\tau^{@v}_j)} \cdot P_{(\mathcal{C},a,m)}^{(\beta,\varepsilon)}((v,\tau^{@v}_j),(w,\tau^{@w}_k)) := \\ 
& & 
\begin{cases} 
1 & \text{if } [v \ne w] \land [\tau^{@v}_j = \tau^{@w}_k] \\ 
\max(\varepsilon, \exp(-\beta[\tau^{@v}_{j+1} - \tau^{@v}_j])) & \text{if } [v = w] \land [j+1 = k] \land [d^+_{(v,\tau^{@v}_{j+1})} > 0] \\ 
\max(\varepsilon, \exp(-\beta[\tau_{(a,m)}^{@v+} - \tau^{@v}_j])) & \text{if } [v = w] \land [j+1 = k] \land [d^+_{(v,\tau^{@v}_{j+1})} = 0] \\ 
1 & \text{if } [v = w] \land [j = k] \land [d^+_{(v,\tau^{@v}_j)} = 0] \\
0 & \text{otherwise}.
\end{cases} \nonumber
\end{eqnarray}
Here $\tau_{(a,m)}^{@v+} := \min(\inf [(a_m,\infty) \cap \mathcal{C}@v], a_M)$, the $Z_{(v,\tau^{@v}_j)}$ are defined so that the rows of $P_{(\mathcal{C},a,m)}^{(\beta,\varepsilon)}$ sum to unity, and $d^+$ denotes outdegree in $T(\mathcal{C})|_{[a_{m-1},a_m)}$. Figure \ref{fig:SimpleMarkov} shows a simple example.

The details of \eqref{eq:transition1} might appear very arbitrary, but upon examination it can be seen that quite the opposite is true. The provision for $\varepsilon > 0$ makes it possible to avoid degeneracies \emph{in silico}
\footnote{
If there are (say) contacts of the form $(v,w,\tau_*)$ and $(w,v,\tau_*)$ with $\tau_* = \tau^{@v}_{|\mathcal{C}@v|-2} = \tau^{@w}_{|\mathcal{C}@w|-2}$, then \eqref{eq:transition1} entails that the probability of a transition from $(v,\tau_*)$ to $(v,a_m)$ or from $(w,\tau_*)$ to $(w,a_m)$ would be exponentially small were it not for the $\varepsilon$ term. While in principle this is not an issue, in numerical practice this leads to floating-point underflow. Taking $\varepsilon > 0$ avoids this problem without significant side effects.
}, 
whereas the absolute requirement $a \cap \tau(\mathcal{C}) = \varnothing$ is imposed to ensure that the Markov chain defined by \eqref{eq:transition1} has exactly $n$ absorbing states, all of the form $(v,a_m)$. Each of these has the corresponding ``emitting'' state $(v,a_{m-1})$, and so it is natural to consider the probability $\mathcal{P}_{(\mathcal{C},a,m)}^{(\beta,\varepsilon)}(v,w)$ of absorption in $(w,a_m)$ when starting from $(v,a_{m-1})$. This quantity can be easily and efficiently computed using the so-called \emph{fundamental matrix} \cite{Bremaud}. Meanwhile, the term $\tau_{(a,m)}^{@v+}$ provides a mechanism to mitigate artificial ``boundary effect'' behavior for $m = M$. Furthermore, \eqref{eq:transition1} leads to a very clean temporal coherence property and a very straightforward physical interpretation.

Regarding the symmetry properties enumerated above, the unit and zero terms in \eqref{eq:transition1} merely express property i), viz. an equivalence among different spatial transitions and the underlying topology of the temporal digraph. Property ii) is embodied in the following 

\begin{proposition}
If $a_0, a_{|a|-1} \in a' \subseteq a$, then
\begin{equation}
\label{eq:DCN2InhomogeneousMarkov}
\mathcal{P}_{(\mathcal{C},a',1)}^{(\beta,\varepsilon)} \cdot \dots \cdot \mathcal{P}_{(\mathcal{C},a',|a'|-1)}^{(\beta,\varepsilon)} = \mathcal{P}_{(\mathcal{C},a,1)}^{(\beta,\varepsilon)} \cdot \dots \cdot \mathcal{P}_{(\mathcal{C},a,|a|-1)}^{(\beta,\varepsilon)}. \ \qed
\end{equation}
\end{proposition}

The import of the proposition is that for any $(\beta,\varepsilon)$ there is a temporally coherent family of time-inhomogeneous Markov chains associated to $\mathcal{C}$. The temporal coherence manifests in two ways: first, the transition probabilities for any given time interval correspond to temporally coherent random paths over that interval; second, transition matrices in successive intervals can be coherently multiplied. 

Note that the proposition does not depend on the specific form of \eqref{eq:transition1}: indeed, the ``$\exp(-\beta \cdot \Delta \tau)$'' terms could be replaced with fairly generic alternatives while still satisfying property ii).
\footnote{
The $\Delta \tau$ dependence is necessary and in the context of information flows is a plausible approximant (for small values) to the conditional Kolmogorov complexity of the intervening computation. 
}
However, these particular terms are required in order to jointly satisfy properties iii) and iv): i.e., memorylessness and self-consistency in the limit $\Delta \tau \downarrow 0$ (the latter of these entails that we cannot multiply the exponentials by some non-unit constant). That is, apart from the numerical underflow-avoiding $\varepsilon$ (which is itself introduced in an obvious way), the form of \eqref{eq:transition1} is completely dictated by the temporal digraph structure in concert with obviously desirable symmetries.
\footnote{
For a weighted DCN, normalizing so that the sum of outbound weights equals either $d^+$ or zero as appropriate and replacing the first case in \eqref{eq:transition1} with the corresponding normalized weight gives an easy and consistent generalization. 
}

The limit $\beta \rightarrow \infty$ evidently amounts to considering so-called \emph{greedy walks} \cite{SaramakiHolme} (more generally, in the regime $\beta > 0$, ``spatial'' transitions are more likely than ``temporal'' transitions), and as we shall see the general construction embeds the temporal coherence of random paths in a more faithful and conservative way than the sorts of series of ``projected snapshots'' that have heretofore characterized attempts such as \cite{PerraEtAl,StarniniEtAl,RochaMasuda,GrindrodHigham} to map a DCN into a time series of graphs and/or provide a substrate for random walks. 

The proposition above provides a mechanism for nominally optimizing the choice of $a$. Write $N := |\mathcal{C}|$, $M := |a|$, and suppose that $a$ is such that $\left |\mathcal{C}|_{[a_m,a_{m+1})} \right | \approx N/M$. Now $|V(T(\mathcal{C}|_{[a_m,a_{m+1})}))| \approx 2(n+N/M)$. Meanwhile, the complexity of computing $\mathcal{P}_{(\mathcal{C},a,j)}^{(\beta,\varepsilon)}$ is dominated by a matrix division step of the form $(I-Q) \backslash R$, where $Q$ is the block of $P_{(\mathcal{C},a,j)}^{(\beta,\varepsilon)}$ whose rows and columns both correspond to transient states, and $R$ is the block whose rows correspond to transient states but whose columns correspond to absorbing states. There are approximately $n+2N/M$ transient states and exactly $n$ absorbing states, so the complexity of computing $(I-Q) \backslash R$ is $O(n(n+2N/M)^{\omega-1})$, where we take matrix multiplication and inversion to have complexity exponent $\omega > 2$ (in practice $\omega = 3$ for dense unstructured matrices). Since there are $M-1$ multiplications, the overall complexity of the RHS of \eqref{eq:DCN2InhomogeneousMarkov} is $O(Mn(n+2N/M)^{\omega-1})$. It is easy to check that $\arg \min_M Mn(n+2N/M)^{\omega-1} = 2(\omega-2)N/n$, and taking this value for $M$ yields a nominal complexity that is linear in $N$. In particular, the only reason to take $M \ll N/n$ is if the complexity of the other operations involved in the overall computation dominates the linear algebra complexity: it is cheaper to invert and multiply a lot of small matrices than to invert and multiply a few large matrices. Taking $M$ to be at least comparable to $N/n$ also has the obvious benefit of providing a more detailed picture of the dynamics of $\mathcal{C}$ than a smaller value. 

Before proceeding further, let us exhibit the basic construction:

\begin{example}
Consider yet again the DCN depicted in figure \ref{fig:DCNtemporalgraph}. Let $\tau_j = j$ for $1 \le j \le 4$, $\varepsilon \ll 1$, $a = \{0,2.5,5\}$ and $a' = \{0,5\}$. We then have that the entries of $P_{(\mathcal{C},a',1)}^{(\beta,\varepsilon)}$ are as shown in figure \ref{fig:SimpleMarkov} and (using $\cdot$ in matrices as shorthand for $0$)
\begin{eqnarray}
\mathcal{P}_{(\mathcal{C},a',1)}^{(\beta,\varepsilon)} & = & \begin{pmatrix}
\frac{e^{\beta}+1}{(e^{4\beta}+1)(e^{\beta}+1)} & \cdot & \frac{e^{5\beta}}{(e^{4\beta}+1)(e^{\beta}+1)} & \frac{e^{4\beta}}{(e^{4\beta}+1)(e^{\beta}+1)} & \cdot \\
\cdot & \frac{1}{e^{2\beta}+1} & \cdot & \cdot & \frac{e^{2\beta}}{e^{2\beta}+1} \\
\cdot & \cdot & 1 & \cdot & \cdot \\
\cdot & \cdot & \frac{e^{\beta}}{e^{\beta}+1} & \frac{1}{e^{\beta}+1} & \cdot \\
\cdot & \cdot & \frac{e^{2\beta}}{(e^{\beta}+1)^2} & \frac{e^{\beta}}{(e^{\beta}+1)^2} & \frac{e^{\beta}+1}{(e^{\beta}+1)^2}
\end{pmatrix} \nonumber \\
& = & 
\begin{pmatrix}
\frac{1}{e^{4\beta}+1} & \cdot & \cdot & \frac{e^{4\beta}}{e^{4\beta}+1} & \cdot \\
\cdot & 1 & \cdot & \cdot & \cdot \\
\cdot & \cdot & 1 & \cdot & \cdot \\
\cdot & \cdot & \cdot & 1 & \cdot \\
\cdot & \cdot & \cdot & \frac{e^{\beta}}{e^{\beta}+1} & \frac{1}{e^{\beta}+1}
\end{pmatrix} 
\cdot 
\begin{pmatrix}
1 & \cdot & \cdot & \cdot & \cdot \\
\cdot & \frac{1}{e^{2\beta}+1} & \cdot & \cdot & \frac{e^{2\beta}}{e^{2\beta}+1} \\
\cdot & \cdot & 1 & \cdot & \cdot \\
\cdot & \cdot & \frac{e^{\beta}}{e^{\beta}+1} & \frac{1}{e^{\beta}+1} & \cdot \\
\cdot & \cdot & \cdot & \cdot & 1
\end{pmatrix} \nonumber \\
& = & \mathcal{P}_{(\mathcal{C},a,1)}^{(\beta,\varepsilon)} \cdot \mathcal{P}_{(\mathcal{C},a,2)}^{(\beta,\varepsilon)}, \nonumber
\end{eqnarray}
so we can see that as $\beta$ increases (equivalently, the temperature decreases) the most likely transitions become those corresponding to temporally coherent paths in which spatial arcs are greedily traversed. Meanwhile, consider the digraph $D$ with edges $(s(c),t(c))$ for $c \in \mathcal{C}$ as well as loops $(v,v)$ for $v \in [n]$: the adjacency matrix of $D$ has nonzero entries in the $(2,3)$ and $(2,4)$ positions as well as in the sparsity pattern of $\mathcal{P}_{(\mathcal{C},a',1)}^{(\beta,\varepsilon)}$. These spurious $(2,3)$ and $(2,4)$ entries correspond to nonexistent temporally coherent paths in $\mathcal{C}$, highlighting that $\mathcal{P}$ gives a more detailed and accurate picture of $\mathcal{C}$ than $D$. 
\qed
\end{example}

\begin{center}
\begin{figure}
	\begin{tikzpicture}[->,>=stealth',shorten >=1pt]
		\node at (-1,-0.6) {$s,t = $ 1};
		\node at (-1,-1.2) {2};
		\node at (-1,-1.8) {3};
		\node at (-1,-2.4) {4};
		\node at (-1,-3.0) {5};
		\node at (0,-3.8) {$\tau = 0$};
		\node at (2,-3.8) {$1$};
		\node at (4,-3.8) {$2$};
		\node at (6,-3.8) {$3$};
		\node at (8,-3.8) {$4$};
		\node at (10,-3.8) {$5$};
		\coordinate (v1a) at (0,-0.6);
		\coordinate (v2a) at (0,-1.2);
		\coordinate (v3a) at (0,-1.8);
		\coordinate (v4a) at (0,-2.4);
		\coordinate (v5a) at (0,-3.0);
		\coordinate (v11) at (2,-0.6);
		\coordinate (v21) at (2,-1.2);
		\coordinate (v31) at (2,-1.8);
		\coordinate (v41) at (2,-2.4);
		\coordinate (v51) at (2,-3.0);
		\coordinate (v42) at (4,-2.4);
		\coordinate (v52) at (4,-3.0);
		\coordinate (v23) at (6,-1.2);
		\coordinate (v53) at (6,-3.0);
		\coordinate (v14) at (8,-0.6);
		\coordinate (v24) at (8,-1.2);
		\coordinate (v34) at (8,-1.8);
		\coordinate (v44) at (8,-2.4);
		\coordinate (v54) at (8,-3.0);
		\coordinate (v1z) at (10,-0.6);
		\coordinate (v2z) at (10,-1.2);
		\coordinate (v3z) at (10,-1.8);
		\coordinate (v4z) at (10,-2.4);
		\coordinate (v5z) at (10,-3.0);
		\foreach \from/\to in {
			v1a/v11, v2a/v23, v23/v2z, v3a/v34, v34/v3z, v4a/v41, v41/v42, v42/v44, v44/v4z, v5a/v52, v52/v53, v53/v5z}
		\draw [dashed] (\from) to (\to);
		\path [->] (v11) edge node [above left] {$Z_{(1,1)}^{-1} \max(\varepsilon, e^{-4\beta})$} (v1z);
		\path [->,out=-60,in=60,looseness=1,gray] (v11) edge node [left] {$Z_{(1,1)}^{-1}$} (v41);
		\path [->] (v52) edge node [below] {$Z_{(5,2)}^{-1} \max(\varepsilon, e^{-\beta})$} (v53);
		\path [->,out=60,in=-60,looseness=1,gray] (v52) edge node [left] {$Z_{(5,2)}^{-1}$} (v42);
		\path [->] (v23) edge node [above] {$Z_{(2,3)}^{-1} \max(\varepsilon, e^{-2\beta})$} (v2z);
		\path [->,out=-60,in=60,looseness=1,gray] (v23) edge node [left] {$Z_{(2,3)}^{-1}$} (v53);
		\path [->] (v44) edge node [below] {$Z_{(4,4)}^{-1} \max(\varepsilon, e^{-\beta})$} (v4z);
		\path [->,out=60,in=-60,looseness=1,gray] (v44) edge node [left] {$Z_{(4,4)}^{-1}$} (v34);
	\end{tikzpicture}
	\caption{\label{fig:SimpleMarkov} Entries of $P_{(\mathcal{C},a',1)}^{(\beta,\varepsilon)}$ not in $\{0,1\}$ for $\mathcal{C} := \{(1,4,1), (5,4,2), (2,5,3), (4,3,4)\}$ and $a' = \{0,5\}$ are indicated along with solid arcs ({\color{gray} gray} for spatial arcs; black for temporal arcs); unit entries correspond to dashed arcs.}
\end{figure}
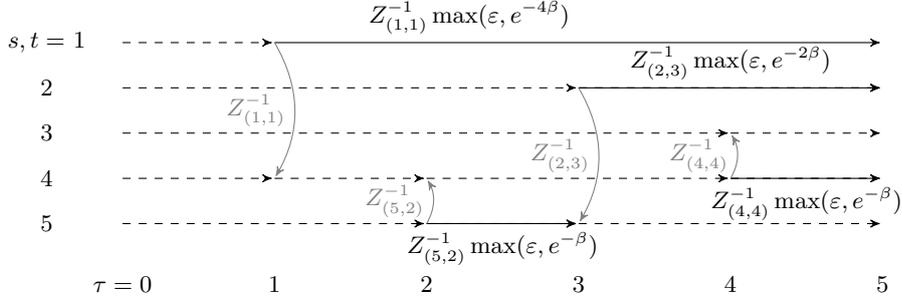
\end{center}

\subsection{\label{sec:Embeddability}Embeddability}

Since a DCN lives in continuous time, it is natural to ask if the Markov chain $\mathcal{P}_{(\mathcal{C},a,m)}^{(\beta,\varepsilon)}$ corresponds to a continuous-time Markov process, i.e., if there is a well-defined notion of instantaneous probability rate for all times. This is an instance of the so-called \emph{Markov embeddability} problem \cite{LencastreEtAl}, and this instance can be answered effectively for various situations of practical interest.

First suppose that no two contacts occur at once. Then the embeddability problem reduces to the case of a single contact for $n=2$, and this in turn follows from the following readily verifiable identity for $p \in (0,1)$:
\begin{equation}
\log \begin{pmatrix} 1-p & p \\ 0 & 1 \end{pmatrix} =  \log (1-p) \cdot \begin{pmatrix} 1 & -1 \\ 0 & 0 \end{pmatrix}. \nonumber
\end{equation}

However, simultaneous contacts can obstruct embeddability. It suffices again to consider the case $n = 2$: here a stochastic matrix is embeddable iff its determinant is positive. But with $\mathcal{C} := \{(1,2,0), (2,1,0), (1,2,\tau), (2,1,\tau)\}$ and $a = \{-1,\tau/2,2 \tau\}$, a quick calculation shows that $\det \mathcal{P}_{(\mathcal{C},a,1)}^{(\beta,0)} < 0$ for $\beta > 0$.

The following result follows immediately from Proposition IV.3 of \cite{LencastreEtAl} and takes both of the preceding cases into account. The algorithm of \S V.B of \cite{LencastreEtAl} can be used to estimate a Markov generator when it exists. 

\begin{proposition}
If $T(\mathcal{C})$ is acyclic, then $\mathcal{P}_{(\mathcal{C},a,m)}^{(\beta,\varepsilon)}$ is embeddable. \qed
\end{proposition}

\section{\label{sec:DRAD}Data reduction and anomaly detection}


In this section, we show how the Markov model performs data reduction well enough to be used as a practical anomaly detector. 

We used data from an adversarial challenge in DARPA's Transparent Computing program. Specifically, we considered a DCN $\mathcal{C}$ formed from $N \approx 3.4 \cdot 10^6$ read and write events spanning a period of four days and derived in turn from data produced by the CADETS tool \cite{JenkinsonEtAl}. We ignored process IDs, resulting in $n = 418$ files (of which 88 were processes). We curated the ground truth data set for this challenge to exclude events that were demonstrably not present in our input data (e.g., directory changes). This left 54 APT events, each given with an accurate timestamp with a precision of 1 second. To translate each APT event record into a form suitable for evaluation, we found the set of contacts that occur within $\pm 1$ second (note that this is conservative), and extracted the subset of these for which either the source or target are a process or file in the APT event record. This yielded a set $\mathcal{G} \subset \mathcal{C}$ of 216 ground truth contacts distributed over the 54 APT events.

We set $\beta$ to the average inter-contact time (as suggested in an earlier footnote), $\varepsilon$ to the square root of machine epsilon, and $a$ to 10-second windows (with $M = 28423$). Writing $\mathcal{P}(m) \equiv \mathcal{P}_{(\mathcal{C},a,m)}^{(\beta,\varepsilon)}$ and $(f \cup g)(X) := f(X) \cup g(X)$, we define for respective probability and relative frequency thresholds $\lambda, \mu \in (0,1)$
\begin{equation}
\label{eq:estimate}
\hat{\mathcal{I}}(m) := (\pi_1 \cup \pi_2) \left ( \left \{ (j,k) : \left [ \mathcal{P}_{jk}(m) > \lambda \right] \land \left [ \left | \{ m' : \mathcal{P}_{jk}(m') > \lambda \} \right |/M < \mu \right ] \right \} \right )
\end{equation}
with $\pi_i$ indicating a projection onto the $i$th factor, and
\begin{equation}
\label{eq:truth}
\mathcal{I}(m) := (s \cup t) \left ( \left \{ c \in \mathcal{G} : \tau(c) \in [a_m,a_{m+1}) \right \} \right )
\end{equation}
That is, $\hat{\mathcal{I}}(m)$ is the set of indices corresponding to origins or destinations of information flows spanning the $m$th time window that are probable (as defined by $\lambda$), and for which this probability is also rare (as defined by $\mu$); meanwhile, $\mathcal{I}(m)$ is the set of indices corresponding to sources or targets of ground truth events during the $m$th time window.
\footnote{
In many cases either the source or the target of a ground truth event does not exist. For example, the userspace commands \texttt{hostname} and \texttt{put /tmp/netrecon} correspond to the $(\text{\emph{process name}},\text{\emph{filename}})$ pairs $(\texttt{hostname},\varnothing)$; and $(\varnothing,\texttt{/tmp/netrecon})$. By way of comparison, the command \texttt{rm -f /tmp/netrecon.log} corresponds to the pair $(\texttt{rm},\texttt{/tmp/netrecon.log})$.
}

\begin{figure}[htbp]
\includegraphics[trim = 10mm 0mm 10mm 0mm, clip, width=60mm,keepaspectratio]{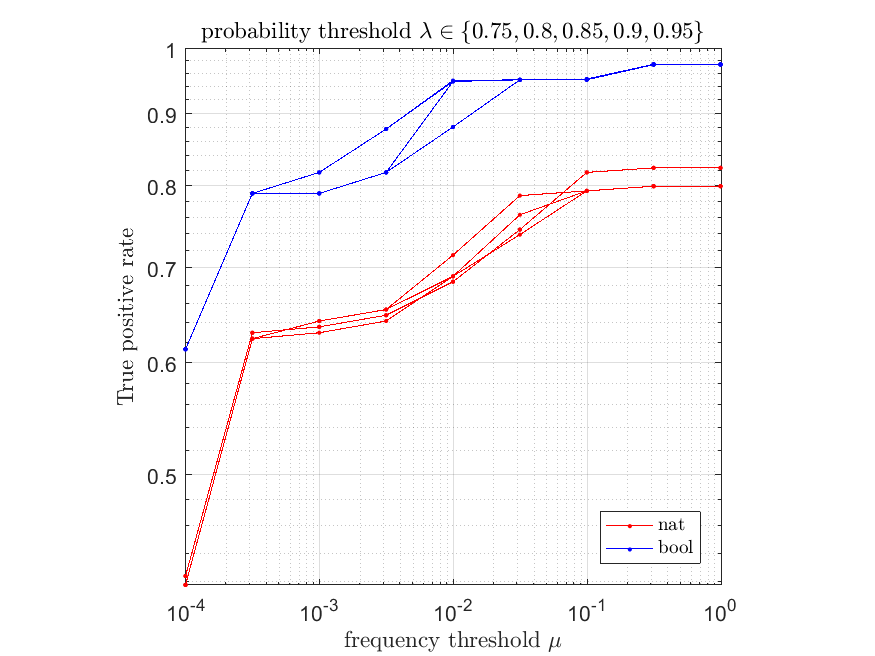}
\includegraphics[trim = 10mm 0mm 10mm 0mm, clip, width=60mm,keepaspectratio]{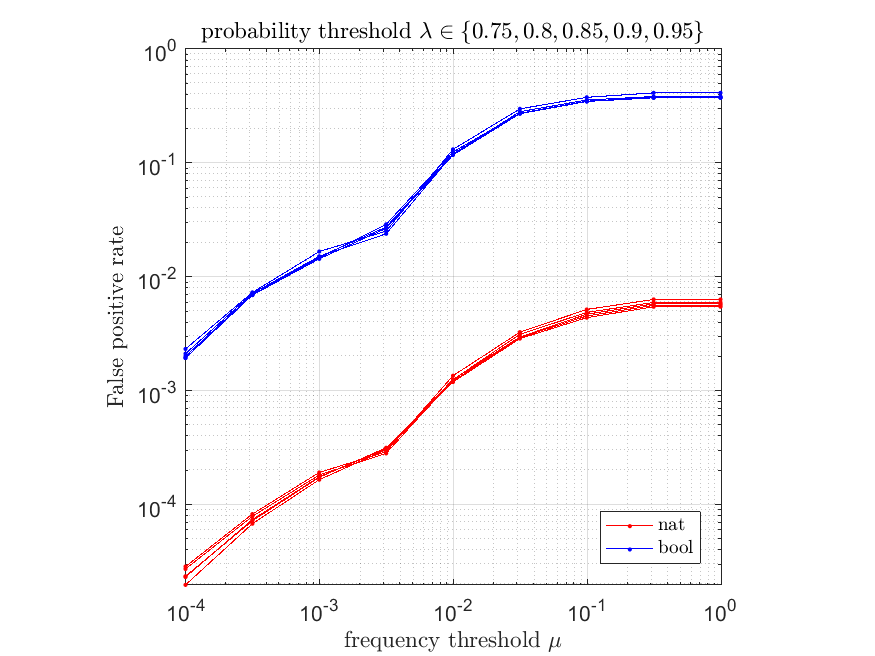}
\caption{ \label{fig:rates} True (left) and false (right) positive rates as defined in \eqref{eq:rates}.} 
\end{figure}

\begin{figure}[htbp]
\includegraphics[trim = 10mm 0mm 10mm 0mm, clip, width=60mm,keepaspectratio]{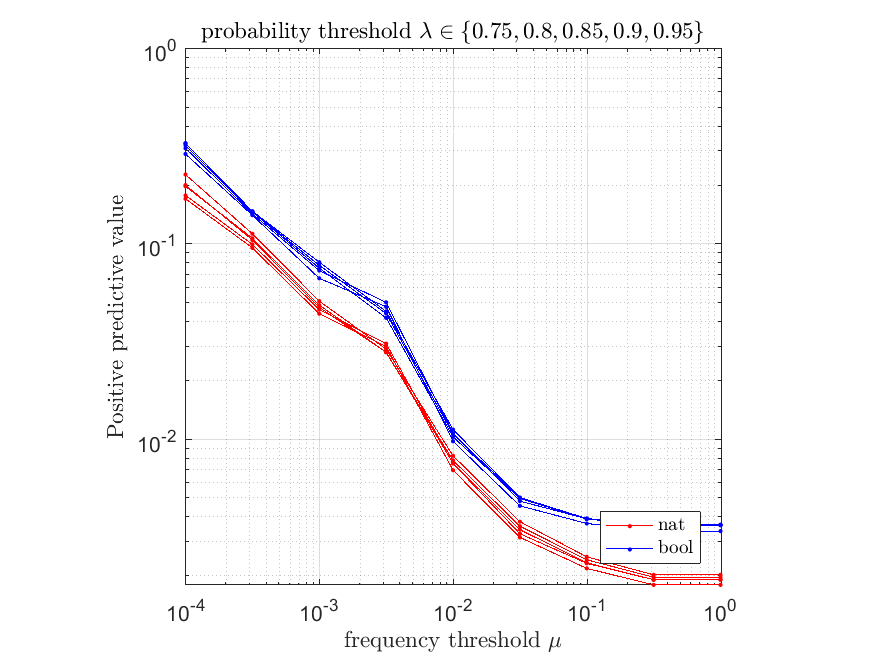}
\includegraphics[trim = 10mm 0mm 10mm 0mm, clip, width=60mm,keepaspectratio]{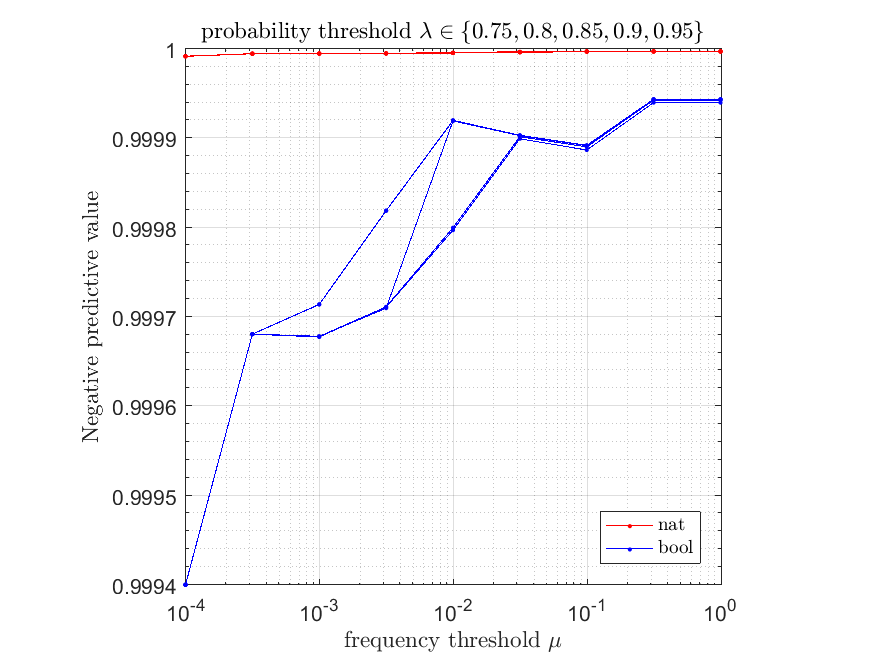}
\caption{ \label{fig:values} Positive (left) and negative (right) predictive values as defined in \eqref{eq:values}.} 
\end{figure}

From \eqref{eq:estimate} and \eqref{eq:truth} we can define both Boolean and natural number versions of detection metrics. The Boolean version uses $\llbracket \top \rrbracket := 1$ and $\llbracket \bot \rrbracket := 0$ \emph{\`a la}
\begin{align}
\label{eq:bool}
\delta^{\top+}_\text{bool}(m) & := & \left \llbracket \left [ \hat{\mathcal{I}}(m) \ne \varnothing \right ] \land \left [ \hat{\mathcal{I}}(m) \cap \mathcal{I}(m) \ne \varnothing \right ] \right \rrbracket; \nonumber \\
\delta^{\bot+}_\text{bool}(m) & := & \left \llbracket \left [ \hat{\mathcal{I}}(m) \ne \varnothing \right ] \land \left [ \hat{\mathcal{I}}(m) \cap \mathcal{I}(m) = \varnothing \right ] \right \rrbracket; \nonumber \\
\delta^{\bot-}_\text{bool}(m) & := & \left \llbracket \left [ \hat{\mathcal{I}}(m) = \varnothing \right ] \land \left [ \mathcal{I}(m) \ne \varnothing \right ] \right \rrbracket; \nonumber \\
\delta^{\top-}_\text{bool}(m) & := & \left \llbracket \left [ \hat{\mathcal{I}}(m) = \varnothing \right ] \land \left [ \mathcal{I}(m) = \varnothing \right ] \right \rrbracket.
\end{align}
The natural number analogues of \eqref{eq:bool} are (suppressing $m$ for brevity)
\begin{equation}
\label{eq:nat}
\delta^{\top+}_\text{nat} := \left | \hat{\mathcal{I}} \cap \mathcal{I} \right |; \quad 
\delta^{\bot+}_\text{nat} := \left | \hat{\mathcal{I}} \cap \mathcal{I}^c \right |; \quad 
\delta^{\bot-}_\text{nat} := \left | \hat{\mathcal{I}}^c \cap \mathcal{I} \right |; \quad 
\delta^{\top-}_\text{nat} := \left | \hat{\mathcal{I}}^c \cap \mathcal{I}^c \right |.
\end{equation}
From these we get in turn the usual detection metrics shown in Figures \ref{fig:rates} and \ref{fig:values}, i.e. true positive rate (or recall) and false positive rate
\begin{equation}
\label{eq:rates}
\text{TPR} := \frac{\sum_m \delta^{\top+}(m)}{\sum_m \delta^{\top+}(m)+\sum_m \delta^{\bot-}(m)}; \quad 
\text{FPR} := \frac{\sum_m \delta^{\bot+}(m)}{\sum_m \delta^{\bot+}(m)+\sum_m \delta^{\top-}(m)},
\end{equation}
and positive predictive value (or precision) and negative predictive value
\begin{equation}
\label{eq:values}
\text{PPV} := \frac{\sum_m \delta^{\top+}(m)}{\sum_m \delta^{\top+}(m)+\sum_m \delta^{\bot+}(m)}; \quad 
\text{NPV} := \frac{\sum_m \delta^{\top-}(m)}{\sum_m \delta^{\bot-}(m)+\sum_m \delta^{\top-}(m)}.
\end{equation}



It is obvious from Figures \ref{fig:rates} and \ref{fig:values} that the results are broadly insensitive to the probability threshold $\lambda$, though it is easy to pick an optimal value using the technique of \cite{Huntsman}. Similarly, a cursory analysis (not shown) indicates that the value of $\beta$ is only important up to several orders of magnitude (this is because potential information flow probabilities strongly tend to be either very near or bounded away from unity in practice, a fact which we also exploit in setting the probability threshold $\lambda$). For the value $\mu = 10^{-3}$, we see that a clear majority of the APT events are detected (by either version of the metrics) with a false positive rate below 2 percent (again, by either version). The results indicate that the model is a sufficiently effective data reduction technique (in particular, the negative predictive value is essentially perfect) to be a useful anomaly detector. 

In fact, of the 57 (out of 418) files which are targets of high-probability potential information flows in the model, 27 fall below the $\mu = 10^{-3}$ level and have backtracks with fewer than 20 (or for that matter, 90) vertices. From these 27, 6 (in 3 pairs) correspond to the 3 executables which the APT wrote to \texttt{/tmp} from its initial foothold.

\section{\label{sec:Remarks}Remarks}

The Markov model depends in an essential way on the detailed structure of the underlying DCN. Experiments not detailed here have shown that inserting even 1\% of random contacts seriously degrades the data reduction and anomaly detection characteristics of the model. This is a good thing, though: it indicates that the model captures the most important aspects of flows from contact data. 

We can also run the model in reverse by simply applying the map $(s,t,\tau) \mapsto (t,s,-\tau)$ beforehand. However, this idea of analyzing reverse information flow remains to be fully explored, as does a related notion of a \emph{taint calculus} proposed by George Cybenko.

An interesting perspective on the model is that it gives us a time-varying geometry. If $d$ denotes an arbitrary metric on probability distributions, we get an induced metric for a given time window of the form $d(v,w) := d(\mathcal{P}_{v\cdot},\mathcal{P}_{w\cdot})$, where $\mathcal{P}_{v\cdot}$ is the $v$th row of $\mathcal{P}$. This in turn allows us to do things like time-dependent clustering. Analysis of the variation of information \cite{Meila} between subsequent clusterings would give additional principled insight into dynamical behavior.

\section*{\label{sec:acknowledgements}Acknowledgements}

The author thanks Yingbo Song, Rob Ross, and Mike Weber for many helpful discussions as well as creating the summary and ground truth data used in \S \ref{sec:DRAD}, and George Cybenko for still more helpful discussions. This material is based upon work supported by the Defense Advanced Research Projects Agency (DARPA) and the Air Force Research Laboratory (AFRL). Any opinions, findings and conclusions or recommendations expressed in this material are those of the author(s) and do not necessarily reflect the views of DARPA or AFRL.

%
%
%
%
%
%
\end{document}